\documentclass[preprint,aps,prd,nofootinbib,superscriptaddress]{revtex4}

\usepackage{graphicx,fancyhdr,lineno,multirow,rotating,calc,amssymb}

\usepackage{epsfig}
\usepackage{epsf}

\def\beq{\begin{equation}}
\def\eeq{\end{equation}}

\def\beqa{\begin{eqnarray}}
\def\eeqa{\end{eqnarray}}

\def\nods#1#2{{#1 \over \sqrt{#2}}}

\def\babar{\mbox{\slshape B\kern-0.1em{\small A}\kern-0.1em
    B\kern-0.1em{\small A\kern-0.2em R}}}

\def\Dz{\ensuremath{D^0}}
\def\piz{\ensuremath{\pi^0}}

\def\ppp{\ensuremath{{\pi^+\pi^-\pi^0}}}

\def\dztoppp{\ensuremath{D^0\to \ppp}}

\def\BR#1{\ensuremath{{\cal B}(#1)}}

\def\ket|#1>{\left|#1 \right>}

\def\bra<#1|{\left< #1 \right|}

\def\bracket<#1|#2>{\setbox0=\vbox{\hbox{$#1$$#2$}}\left<#1\kern1pt \vrule  height\ht0\kern2pt #2\right>}

\def\matrel<#1|#2|#3>{\setbox0=\vbox{\hbox{$#1$$#2$$#3$}}\left<#1\kern1pt \vrule height\ht0\kern1pt#2\kern1pt \vrule height\ht0\kern1pt #3\right>}

\def\Ipair{\ensuremath{I_{12}}}
\def\sp{\ensuremath{s_+}}
\def\sm{\ensuremath{s_-}}

\begin{document}

\title{{\large \bf Isospin analysis of $D^0$ decay to three pions}}
\bigskip

\author{M. Gaspero}
\affiliation{Universit\`a di Roma La Sapienza, Dipartimento di Fisica and INFN, I-00185 Roma, Italy }
\author{B. Meadows}
\affiliation{University of Cincinnati, Cincinnati, Ohio 45221, USA}
\author{K. Mishra}
\altaffiliation{Now at Fermi National Accelerator Laboratory, Batavia, IL, USA}
\affiliation{University of Cincinnati, Cincinnati, Ohio 45221, USA}
\author{A. Soffer}
\affiliation{Tel Aviv University, Tel Aviv, 69978, Israel}

\date{\today}

\bigskip
\bigskip

\begin{abstract}

The final state of the decay \dztoppp\ is analyzed in terms of isospin
eigenstates. It is shown that the final state is dominated by the
isospin-0 component. This suggests that isospin considerations may
provide insight into this and perhaps other $\Dz$-meson decay. We also discuss 
the isospin nature of the nonresonant contribution in the decay,
which can be further understood by studying the decay
$\Dz\to\piz\piz\piz$.

\end{abstract}
\maketitle

\bigskip
\bigskip

\section{Introduction}
\begin{figure}[!htbp]
  \begin{center}
\includegraphics[width=0.6\textwidth]{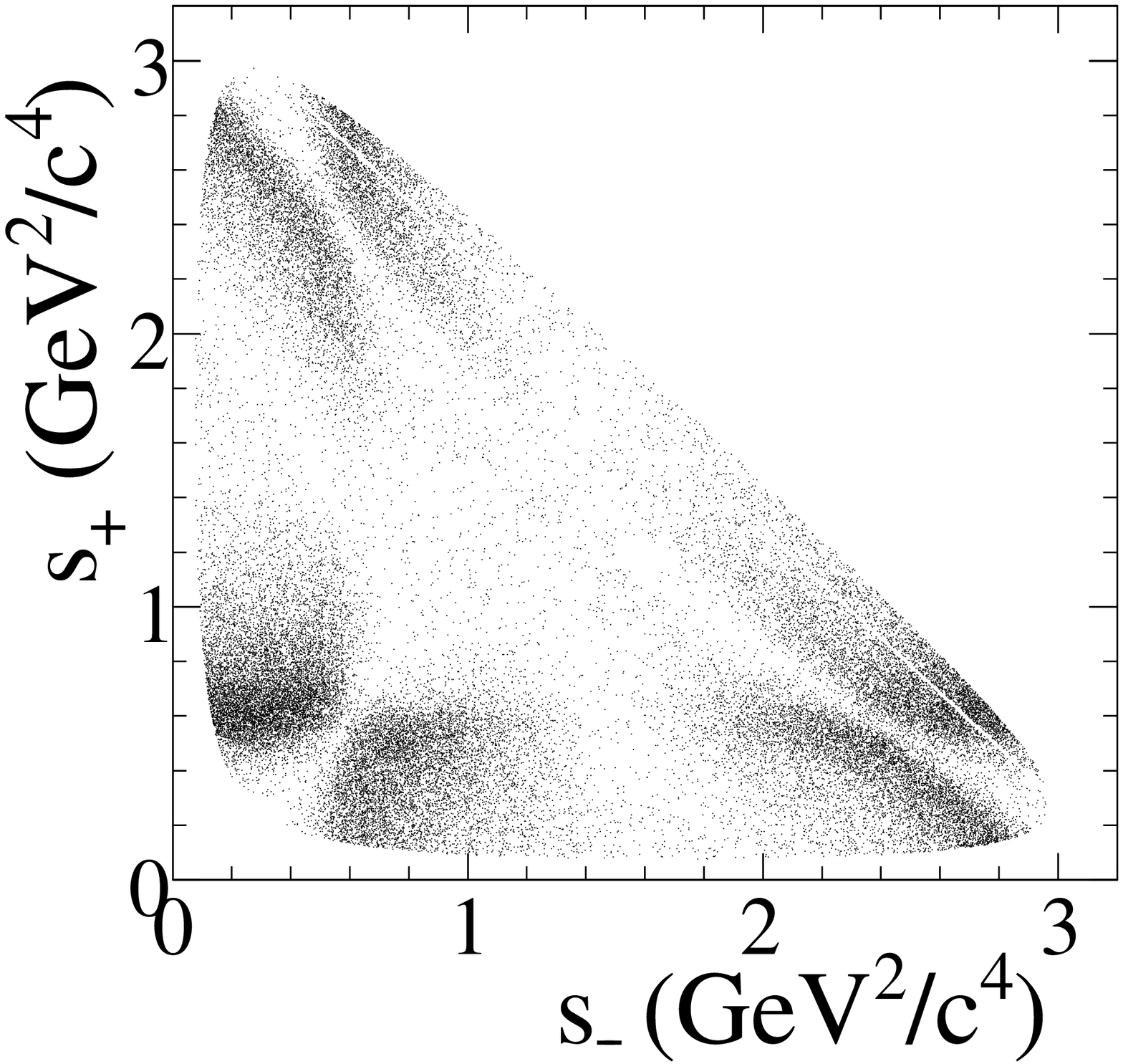}
\caption{The Dalitz-plot distribution of \dztoppp\ events,
from Ref.~\cite{Aubert:2007ii}.}
\label{fig:DP}
\end{center}
\end{figure}

An analysis of the resonant sub-structure in the decay \dztoppp\ was
recently performed by the \babar\ collaboration~\cite{Aubert:2007ii}.
The Dalitz-plot distribution of the \dztoppp\ events
(Fig.~\ref{fig:DP}) shows a clear six-fold symmetry, with the
probability density function vanishing along three axes. As first
described by Zemach~\cite{ref:zemach} and noted in
Ref.~\cite{Aubert:2007ii}, this behavior is indicative of a final
state with isospin $I=0$.

In the \babar\ analysis, the Dalitz-plot distribution is described 
by a probability density function
formed from a wave function taken to be the sum of $N_r$
contributions,
\beq
\psi(\sp, \sm) = \sum_{r}^{N_r} B_r\, g_r(\sp,\sm),
\label{eq:psi}
\eeq
where
$\sp \equiv (p_{\pi^+} + p_{\pi^0})^2$ and 
$\sm \equiv (p_{\pi^-} + p_{\pi^0})^2$  are the squared 
invariant masses of the 
$\pi^+\pi^0$ and $\pi^-\pi^0$ pairs, respectively, 
$B_r$ is a complex coefficient, 
and $g_r(\sp,\sm)$ is the distribution of contribution $r$,
whose functional form is outlined in Ref.~~\cite{Aubert:2007ii}.
The definitions of $g_r(\sp,\sm)$ used here differ
from that of Ref.~\cite{Aubert:2007ii}, in that we define these 
functions to be normalized over the Dalitz plot,
\beq
\int d\sp d\sm \left|g_r(\sp,\sm)\right|^2 = 1.
\label{eq:norm1}
\eeq
The values for the $B_r$ coefficients consistent with
Eqs.~(\ref{eq:psi}) and~ (\ref{eq:norm1}) are reproduced in
Table~\ref{tab:DalitzPlotFit}.

\begin{table}[htbp]
\caption{\label{tab:DalitzPlotFit} Amplitude coefficients
$B_r =  \left|B_r\right|e^{i\phi_r}$ of the contributing final states of
the decay $D^0\to \pi^-\pi^+\pi^0$, adapted from Ref.~\cite{Aubert:2007ii}.
The $f_0(400)$ was labeled $\sigma(400)$ in Ref.~\cite{Aubert:2007ii}.
}
\centering
\begin{tabular}{l|l|l}
\hline\hline
Final state $r$ & Amplitude $\left|B_r\right|$ & Phase $\phi_r$ ($^\circ$) \cr
\hline
Nonresonant   & 0.106 $\pm$ 0.013 $\pm$ 0.014 & $-$11$\pm$4$\pm$2\cr  
\hline
$\rho(770)^+\pi^-$ & 1                         &  0.0 \cr
$\rho(770)^0\pi^0$ & 0.588 $\pm$ 0.006 $\pm$ 0.002 &  16.2$\pm$0.6$\pm$0.4\cr
$\rho(770)^-\pi^+$ & 0.714 $\pm$ 0.008 $\pm$ 0.002 & $-$2.0$\pm$0.6$\pm$0.6\cr
\hline
$\rho(1450)^+\pi^-$& 0.040 $\pm$ 0.011 $\pm$ 0.024 &  $-$146$\pm$18$\pm$24\cr
$\rho(1450)^0\pi^0$& 0.062 $\pm$ 0.012 $\pm$ 0.007 &  10$\pm$8$\pm$13 \cr
$\rho(1450)^-\pi^+$& 0.154 $\pm$ 0.010 $\pm$ 0.007 &  16$\pm$3$\pm$3 \cr
\hline
$\rho(1700)^+\pi^-$& 0.236 $\pm$ 0.019 $\pm$ 0.014 & $-$17$\pm$2$\pm$3 \cr
$\rho(1700)^0\pi^0$& 0.267 $\pm$ 0.016 $\pm$ 0.014 & $-$17$\pm$2$\pm$2 \cr
$\rho(1700)^-\pi^+$& 0.210 $\pm$ 0.012 $\pm$ 0.007 & $-$50$\pm$3$\pm$3 \cr
\hline
$f_0(980)\pi^0$    & 0.056 $\pm$ 0.005 $\pm$ 0.006 & $-$59$\pm$5$\pm$4 \cr
$f_0(1370)\pi^0$   & 0.072 $\pm$ 0.010 $\pm$ 0.010 & 156$\pm$9$\pm$6 \cr
$f_0(1500)\pi^0$   & 0.074 $\pm$ 0.007 $\pm$ 0.007 & 12$\pm$9$\pm$4 \cr 
$f_0(1710)\pi^0$   & 0.072 $\pm$ 0.010 $\pm$ 0.011 & 51$\pm$8$\pm$7 \cr
$f_2(1270)\pi^0$   & 0.130 $\pm$ 0.005 $\pm$ 0.026 & $-$171$\pm$3$\pm$4\cr 
$f_0(400)\pi^0$ & 0.104 $\pm$ 0.008 $\pm$ 0.017 & 8$\pm$4$\pm$8 \cr
\hline\hline
\end{tabular}
\end{table}

The goal of this paper is to quantify the extent to which the $I=0$
component dominates the final state and learn about the contributions
of the other isospin eigenstates.  
In Section~\ref{sec:isospin} we perform an isospin analysis of the
\ppp\ final state. The observed dominance of the $I=0$ component
suggests that isospin considerations are more useful for developing
an understanding of this decay. In Section~\ref{sec:discussion}
we discuss our results, the nature of the nonresonant contribution 
to the decay, a possible mechanism for the observed 
$I=0$ dominance, and further measurements that will help clarify 
outstanding questions.

\section{Isospin Decomposition}
\label{sec:isospin}

Next, we analyze the decay \dztoppp\ in terms of isospin eigenstates.
The 3-pion final state can be described in terms of the total isospin
$I$, the isospin $\Ipair$ of two of the three pions, and the
$z$-projection $I^z$, which is always $0$ for this final state. 
The seven eigenstates $\ket|I(\Ipair)>$ of these quantum numbers that
also satisfy $I^z = 0$ can be written as a linear combination of the
three-pion final states using the appropriate Clebsh-Gordan
coefficients:
\beqa
\ket|3(2)> &=& {1 \over \sqrt{10}} \Bigl(\ket|+0-> + \ket|0+->
                                           + \ket|+-0> + \ket|-+0>
                                           + \ket|0-+> + \ket|-0+>
					   + 2\ket|000>\Bigr)
   \nonumber\\
\ket|2(2)> &=& {1 \over 2} \Bigl(\ket|+0-> + \ket|0+-> 
                                          - \ket|0-+> - \ket|-0+>\Bigr),
   \nonumber\\
\ket|1(2)> &=& {1\over\sqrt{60}}\left[ 3\Bigl(\ket|+0-> + \ket|0+->
                                              +\ket|0-+> + \ket|-0+>\Bigr)
             - 2\Bigl(\ket|+-0> + \ket|-+0>\Bigr) - 4\ket|000>\right],
   \nonumber\\
\ket|2(1)>&=& {1 \over \sqrt{12}} \left[ \ket|+0-> - \ket|0+->
                                    + 2 \Bigl(\ket|+-0> - \ket|-+0>\Bigr)
                                      + \ket|0-+> - \ket|-0+>
				       \right],
   \nonumber\\
\ket|1(1)> &=& {1 \over 2} \Bigl( 
                          \ket|+0-> - \ket|0+-> - \ket|0-+> +\ket|-0+>
                                     \Bigr),
   \nonumber\\
\ket|0(1)> &=& {1\over \sqrt{6}}\Bigl(  \ket|+0-> - \ket|0+->
                                          - \ket|+-0> + \ket|-+0>
                                          + \ket|0-+> - \ket|-0+> \Bigr),
   \nonumber\\
\ket|1(0)> &=& {1 \over \sqrt{3}} \Bigl(\ket|+-0> - \ket|000> + \ket|-+0>
                                      \Bigr),
\label{eq:triplet}
\eeqa
where we have used the notation 
\beqa
\ket|+0-> &=& \ket|1,1> \ket|1,0> \ket|1,-1> = 
  \ket|\pi^+> \ket|\pi^0> \ket|\pi^->, \nonumber\\ 
\ket|000> &=& \ket|1,0> \ket|1,0> \ket|1,0> = 
  \ket|\pi^0> \ket|\pi^0> \ket|\pi^0>,
\eeqa
etc., and it is implied that the first two pions are in an isospin 
eigenstate whose eigenvalue is indicated by the bracketed number $\Ipair$.

The three states in Eq.~(\ref{eq:triplet}) for which $\Ipair=1$ are
identified as those with a $\rho(770)$, $\rho(1450)$, or $\rho(1700)$.
We denote these states as $\rho_n \pi$ according to their radial
excitation quantum number $n \in \{1,2,3\}$, and use $\rho^+$,
$\rho^0$, and $\rho^-$ to indicate any linear combination of these
states with specific electric charge.  We define the $\rho$ states to be
\beqa
\ket|\rho^+> &=& \ket|1,1> 
    = {1 \over \sqrt{2}} \Bigl(\ket|+0> - \ket|0+> \Bigr),
   \nonumber\\
\ket|\rho^0> &=& -\ket|1,0> 
    = {1 \over \sqrt{2}} \Bigl(\ket|-+> - \ket|+->  \Bigr),
   \nonumber\\
\ket|\rho^-> &=& \ket|1,-1> 
    = {1 \over \sqrt{2}} \Bigl(\ket|0-> - \ket|-0>  \Bigr) ,
\label{eq:rhos}
\eeqa
where the minus sign in the $\ket|\rho^0>$ definition 
implies that there is no sign change under cyclic permutations
of the three pions, maintaining consistency with the definitions
used in Ref.~\cite{Aubert:2007ii}.
Given Eq.~(\ref{eq:rhos}), the $\Ipair=1$ states in
Eq.~(\ref{eq:triplet}) can be written as
\beqa
\ket|2(1)>&=& {1 \over \sqrt{6}} \Bigl( \ket|\rho^+\pi^-> 
                                       -2 \ket|\rho^0\pi^0>
                                        + \ket|\rho^-\pi^+>\Bigr), 
\nonumber\\
\ket|1(1)>&=& {1 \over \sqrt{2}} \Bigl( \ket|\rho^+\pi^-> 
                                        - \ket|\rho^-\pi^+>\Bigr), 
\nonumber\\
\ket|0(1)>&=& {1 \over \sqrt{3}} \Bigl( \ket|\rho^+\pi^-> 
                                        + \ket|\rho^0\pi^0>
                                        + \ket|\rho^-\pi^+> \Bigr),
\label{eq:I12=1-states}
\eeqa
where the sign of each $\ket|\rho\pi>$ state is such that it is
symmetric under cyclic permutations of the three pions and
anti-symmetric under the exchange of any pair of pions.

The $\pi^+\pi^-\pi^0$ part of the state $\ket|1(0)>$ is identified as
the sum of the contributions involving the two-body, $I=0$
resonances $f_i$, with $i=0,2$. We therefore write
\beq
\ket|1(0)> = {1 \over \sqrt{3}} \Bigl(\sqrt{2} \ket|f\pi^0> 
                                            - \ket|000>  \Bigr).
\label{eq:100}
\eeq

Since there are no $I=2$ resonances in
Table~\ref{tab:DalitzPlotFit}, the $\Ipair=2$ states in
Eq.~(\ref{eq:triplet}) have no resonant contributions.  However, the
symmetry of the $\pi^+\pi^-\pi^0$ components of $\ket|3(2)>$ indicates
that it may be identified with the nonresonant contribution of
Table~\ref{tab:DalitzPlotFit}. Alternatively, it may constitute the
$\pi^+\pi^-\pi^0$ component of the symmetric $I=1$ state
\beqa
\ket|1(S)> &\equiv& {2 \over 3} \ket|1(2)> + {\sqrt{5} \over 3} \ket|1(0)>
  \nonumber\\
&=& {1 \over \sqrt{15}} \Bigl(\ket|+0-> + \ket|0+->
                           + \ket|+-0> + \ket|-+0>
                           + \ket|0-+> + \ket|-0+>
			   - 3\ket|000>\Bigr).
\label{eq:1S}
\eeqa
In principle, the observed nonresonant state may be a superposition of
$\ket|1(S)>$ and $\ket|3(2)>$. However, the $\ket|1(S)>$ state is
expected to dominate, due to the following argument. 
The four-quark final state produced by the weak decay $c\to d
\overline d u$, shown in Fig.~\ref{fig:diagrams}, cannot have
$I=3$. Since production of the third $q\overline q$ pair will be dominated
by the strong-interaction, it will not change the total isospin. 
Therefore, $I=3$ is disfavored.
It is also possible that a very broad, $\pi^+\pi^-$ $S$-wave resonance is
present in these decays, and that it was partly described by the
constant nonresonant term in the fit in Ref.~\cite{Aubert:2007ii}. In
that case, it would contribute only to the $\ket|1(0)>$ isospin
eigenstate.

In what follows, we take the nonresonant contribution $\ket|NR>$ to be
due only to $\ket|1(S)>$. Then Eqs.~(\ref{eq:100}) and~(\ref{eq:1S})
yield the relation
\beq
\ket|1(2)> = {3 \over \sqrt{10}} \ket|NR> - \sqrt{5 \over 6} \ket|f\pi^0>
   - {2 \over \sqrt{15}} \ket|000>.
\label{eq:120}
\eeq

%
%

We now reorder the terms of Eq.~(\ref{eq:psi})
according to their $\Ipair$ eigenvalues:
\beqa
\psi(\sp, \sm) &=& B_{\rm NR}\, g_{\rm NR}(\sp,\sm) \nonumber\\ 
       &+& B_{\rho^+\pi^-} \, g_{\rho^+\pi^-}(\sp,\sm) \nonumber\\
       &+& B_{\rho^0\pi^0} \, g_{\rho^0\pi^0}(\sp,\sm) \nonumber\\
       &+& B_{\rho^-\pi^+} \, g_{\rho^-\pi^+}(\sp,\sm) \nonumber\\
       &+& B_{f\pi^0} \, g_{f\pi^0}(\sp,\sm),
\label{eq:psi-reorg}
\eeqa
where the first term is the nonresonant term, the last is a sum
over the six final states with $\Ipair=0$ resonances listed at the
bottom of Table~\ref{tab:DalitzPlotFit}, and each of the second, third,
and fourth terms is a sum over the three $\Ipair=1$ $\rho\pi$ states.
For example, 
\beq
g_{\rho^+\pi^-}(\sp,\sm) \equiv 
  {S_{\rho^+\pi^-} \over 
  N_{\rho^+\pi^-}}
\, \exp\left[-i\delta_{\rho^+\pi^-}\right],
\label{eq:norm-wf-def}
\eeq
where
\beqa
S_{\rho^+\pi^-} &\equiv&   
  \sum_{n=1}^3 B_{\rho_{n}^+\pi^-} \, g_{\rho_{n}^+\pi^-}(\sp,\sm),
\nonumber\\
\delta_{\rho^+\pi^-} &\equiv& \arg\left(S_{\rho^+\pi^-}\right),
\nonumber\\
N_{\rho^+\pi^-} &\equiv& \sqrt{\int d\sp d\sm \left|S_{\rho^+\pi^-}\right|^2},
\label{eq:sum-states-def}
\eeqa
and $\rho_n$ ($n=1,2,3$) indicates the three $\rho$ resonances of
Table~\ref{tab:DalitzPlotFit}.  
With these definitions, the wave
function $g_{\rho^+\pi^-}(\sp,\sm)$ is explicitly normalized and has
vanishing average phase.
Requiring that Eq.~(\ref{eq:psi-reorg}) be
identical to~(\ref{eq:psi}) leads to the following values for the
coefficients of Eq.~(\ref{eq:psi-reorg}):
\beqa
B_{\rm NR} &=& 0.1066
\,  e^{-i\, 11.4^\circ}, 
    \nonumber\\
B_{\rho^+\pi^-}   &\equiv&
N_{\rho^+\pi^-}
 \,  \exp\left[i\delta_{\rho^+\pi^-}\right] 
= 1.1976  \,  e^{-i\, 4.3^\circ}, 
    \nonumber\\
B_{\rho^0\pi^0}   &\equiv&
N_{\rho^0\pi^0}
 \,  \exp\left[i\delta_{\rho^0\pi^0}\right] 
= 0.8867 \,  e^{i\, 6.3^\circ}, 
    \nonumber\\
B_{\rho^-\pi^+}   &\equiv&
N_{\rho^-\pi^+}
 \,  \exp\left[i\delta_{\rho^-\pi^+}\right] 
= 1.0077 \,  e^{-i\, 8.2^\circ}, 
    \nonumber\\
B_{f\pi^0}   &\equiv&
N_{f\pi^0}  
 \,  \exp\left[i\delta_{f\pi^0}\right] 
= 0.0700 \,  e^{i\, 40.0^\circ}, 
\label{eq:sum-coeff}
\eeqa
where the symbols $N_s$ and $\delta_s$ for final state $s$ are 
defined analogously to Eq.~(\ref{eq:sum-states-def}).
The value of $B_{\rm NR}$ is taken from
Table~\ref{tab:DalitzPlotFit} and the rest are calculated
numerically as in Eqs.~(\ref{eq:norm-wf-def}) and~(\ref{eq:sum-states-def}).
The phase convention is that of Table~\ref{tab:DalitzPlotFit},
namely, $\delta_{\rho_1^+\pi^-} \equiv 0$.

Next, we write the wave function of Eq.~(\ref{eq:psi-reorg})
as a sum over the Dalitz-plot representations of the eigenstates of
$I$ and $\Ipair$ of Eq.~(\ref{eq:triplet}):
\beqa 
\psi(\sp, \sm) &=& C_{1(2)}\, M_{1(2)}(\sp,\sm)  \nonumber\\
               &+& C_{2(1)}\, M_{2(1)}(\sp,\sm)  \nonumber\\
               &+& C_{1(1)}\, M_{1(1)}(\sp,\sm)  \nonumber\\
               &+& C_{0(1)}\, M_{0(1)}(\sp,\sm)  \nonumber\\
               &+& C_{1(0)}\, M_{1(0)}(\sp,\sm), 
\label{eq:isospin-breakdown}
\eeqa
where $M_{I(\Ipair)}(\sp,\sm)$ is the normalized distribution function
of the eigenstate $\ket|I(\Ipair)>$, obtained by linearly combining
the functions $g_x(\sp,\sm)$ of Eq.~(\ref{eq:psi-reorg}) with the
coefficients of either Eq.~(\ref{eq:I12=1-states}), (\ref{eq:100}),
or~(\ref{eq:120}).
Terms for $\ket|3(2)>$ and $\ket|2(2)>$ were not included in
Eq.~(\ref{eq:isospin-breakdown}), as reasoned earlier.
Then from the definition of $M_{I(\Ipair)}(\sp,\sm)$ follows the desired
transformation between the resonance-based fit coefficients and the
isospin coefficients:
\beqa
C_{1(2)}    &=&  {\sqrt{10} \over 3}\, B_{\rm NR}  , \nonumber\\
C_{2(1)}    &=&  \nods16 \left( B_{\rho^+\pi^-} 
                  -2 B_{\rho^0\pi^0} 
                  + B_{\rho^-\pi^+}\right)  , \nonumber\\
C_{1(1)}    &=&  \nods12 \left( B_{\rho^+\pi^-} 
                  - B_{\rho^-\pi^+}\right)  , \nonumber\\
C_{0(1)}    &=&  \nods13 \left( B_{\rho^+\pi^-} 
                  + B_{\rho^0\pi^0} 
                  + B_{\rho^-\pi^+}\right)  , \nonumber\\
C_{1(0)}    &=&  \sqrt{3 \over 2} \, B_{f\pi^0} + \sqrt{5 \over 6} C_{1(2)},
\label{eq:C-expressions}
\eeqa
where the expressions for $C_{1(0)}$ and $C_{1(2)}$ were chosen so as
to satisfy the $\pi^+\pi^-\pi^0$ projection of Eqs.~(\ref{eq:100})
and~(\ref{eq:120}).

Taking the numerical values of the $B_r$ coefficients from
Eq.~(\ref{eq:sum-coeff}) and Table~\ref{tab:DalitzPlotFit},
Eq.~(\ref{eq:C-expressions}) gives
\beqa
%
%
C_{1(2)}    &=&   (0.0629  \pm  0.0028)\, 
   \exp\left[i \, (-8.9 \pm 2.6)^\circ\right] , \nonumber\\
C_{2(1)}    &=&  (0.1395  \pm 0.0016) \, 
   \exp\left[i \, (-42.5 \pm 0.7)^\circ\right] , \nonumber\\
C_{1(1)}    &=&  (0.0814  \pm 0.0023) \, 
   \exp\left[i \, (18.0 \pm 2.0)^\circ\right] , \nonumber\\
C_{0(1)}    &\equiv&  1, \nonumber\\
C_{1(0)}    &=&  (0.0954 \pm  0.0052) \, 
   \exp\left[i \, (14.5 \pm 2.4)^\circ\right] ,
%
%
\label{eq:C-values-norm}
\eeqa
where we have normalized the coefficients so that $C_{0(1)} = 1$.
The errors reflect the full error matrix of the results presented in
Table~\ref{tab:DalitzPlotFit}~\cite{ref:thesis}.
The correlation matrix for these coefficients are given in
Table~\ref{tab:errMatrix}. 

Eq.~(\ref{eq:C-values-norm}) quantifies the observation, made
qualitatively in Ref.~\cite{Aubert:2007ii} on the basis of the
symmetry exhibited by the Dalitz-plot distribution, that the final
state of the decay \dztoppp\ is dominated by an $I=0$ component.

\begin{table}[htbp]
  \caption{\label{tab:errMatrix} Correlation matrix for the $C_{I(I12)}$ 
    amplitude coefficients of Eq.~(\ref{eq:C-values-norm}).}
  \centering
  \begin{tabular}{l|r|r|r|r|r|r|r|r}
    \hline\hline
                        & $|C_{1(2)}|$ & $\arg(C_{1(2)})|$ & $|C_{2(1)}|$ & $\arg(C_{2(1)})$  &  $|C_{1(1)}|$  & $\arg(C_{1(1)})$  & $|C_{1(0)}|$ & $\arg(C_{1(0)})$\\
    \hline
    $|C_{1(2)}|$&        1             & $-0.120$          & 0.105        &  $-0.018$         &  0.631         & 0.110             & 0.279        & 0.657 \\
    $\arg(C_{1(2)})$&    $-0.120$      & 1                 & 0.062        & 0.106             & $-0.211$       & 0.539             & $-0.760$     & 0.136 \\
    $|C_{2(1)}|$&        0.105         & 0.062             & 1            & 0.008             & 0.179          & 0.029             & $-0.017$     & 0.078 \\
    $\arg(C_{2(1)})$&    $-0.018$      & 0.106             & 0.008        & 1                 & 0.148          & 0.333             & 0.110        & 0.151 \\
    $|C_{1(1)}|$&        0.631         & $-0.211$          & 0.179        & 0.148             & 1              & 0.050             & 0.259        & 0.288 \\
    $\arg(C_{1(1)})$&    0.110         & 0.539             & 0.029        & 0.333             & 0.050          & 1                 & $-0.296$     & 0.097 \\
    $|C_{1(0)}|$&        0.279         & $-0.760$          & $-0.017$     & 0.110             & 0.259          & $-0.296$          & 1            & 0.077 \\
    $\arg(C_{1(0)})$&    0.657         & 0.136             & 0.078        & 0.151             & 0.288          & 0.097             &0.077      & 1 \\
    \hline\hline
  \end{tabular}
\end{table}
\section{Discussion and Conclusions}
\label{sec:discussion}

\begin{figure}
\begin{center}
\includegraphics[width=0.5\textwidth]{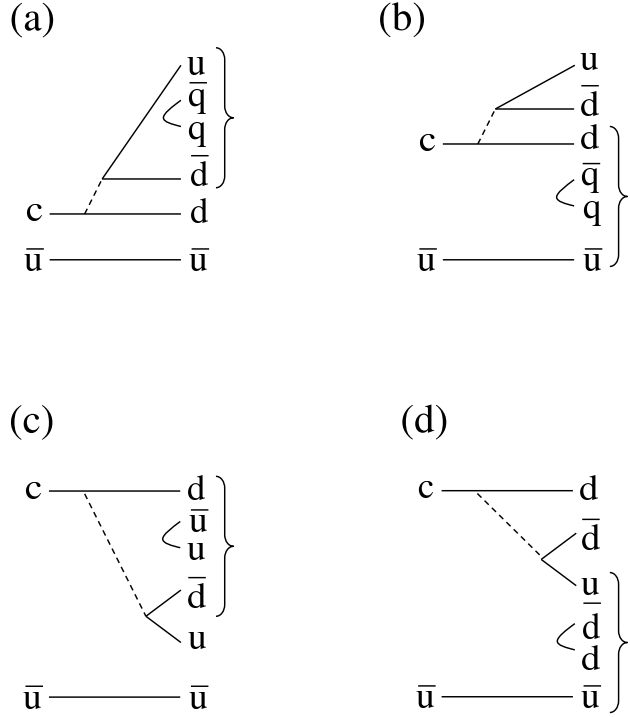}
\caption{\label{fig:diagrams}Feynman diagrams for the decay
  \dztoppp. With curly brackets indicating
   a resonance, the diagrams correspond to the decays
  (a) $D^0\to\rho^+\pi^-$, (b) $D^0\to\pi^+\rho^-$,
  and (c,d) $D^0\to\rho^0\pi^0$ or $D^0\to f\pi^0$.}
\end{center}
\end{figure}

We have analyzed the relative contributions of different components to
the decay \dztoppp\ using results published by
\babar~\cite{Aubert:2007ii}. It appears that isospin considerations
may form a solid basis for understanding the observed decay pattern,
as the amplitude of the $\ket|0(1)>$ final state dominates by factors
of seven or more over the other isospin components.
This dominance has no natural explanation in the decay mechanisms
suggested by the factorization-motivated diagrams of this decay, shown
in Fig.~\ref{fig:diagrams}. While factorization is useful in
predicting the behavior of $B$-meson decays, it is not as
successful when applied to the lighter $D$ mesons.  
The observed $\ket|0(1)>$ dominance in the decay \dztoppp\ may lead to
a better general understanding of charmed meson decays.
Alternatively, perhaps the $I=0$ component is enhanced by the presence
of a yet-unknown and possibly broad state with this quantum number,
which couples strongly to three pions. An inclusive search for such a
state may answer this question.

In conducting the isospin analysis, we took only the $\pi^+\pi^-\pi^0$
projections of the isospin-eigenstates $\ket|1(2)>$ and $\ket|1(0)>$.
The CLEO collaboration~\cite{Rubin:2005py} has set an upper limit of
$3.4\times 10^{-4}$ on the branching fraction $\BR{D^0\to
\pi^0\pi^0\pi^0}$.
Together with the \babar~\cite{ref:dppp} measurement of $\BR{\dztoppp}
= (1.493 \pm 0.057)\%$, this implies an upper limit on the amplitude
ratio ${A(D^0\to \pi^0\pi^0\pi^0) / A(\dztoppp)} < 0.15$,
consistent with the suppression seen in the coefficients $C_{1(2)}$
and $C_{1(0)}$, and the expectation from Eqs.~(\ref{eq:100})
and~(\ref{eq:120}).

As discussed above, the $\pi^+\pi^-\pi^0$ nonresonant amplitude may be
a combination of $\ket|3(2)>$, $\ket|1(S)>$, and a broad $\pi^+\pi^-$
resonance term in $\ket|1(0)>$. 
If it is due only to the $\ket|3(2)>$, Eq.~(\ref{eq:triplet})
predicts the ratio between the
nonresonant $\pi^0\pi^0\pi^0$ and $\pi^+\pi^-\pi^0$ amplitudes to
be $R_{NR} = \sqrt{2/3}$. By contrast, $\ket|1(S)>$-dominance leads to
$R_{NR} = \sqrt{3/2}$, from Eq.~(\ref{eq:1S}).
In the $\ket|1(0)>$ case, the ratio between the nonresonant
$\pi^0\pi^0\pi^0$ amplitude and the sum of the $f\pi^0$ and nonresonant
$\pi^+\pi^-\pi^0$ amplitudes should be $1/\sqrt{2}$.
We note that the ratio $R_{NR} = \sqrt{1.556 \pm 0.012}$ is observed
in $K_L$ decays to three pions, where the nonresonant contribution accounts
for over 95\% of the branching fractions.  The same 
situation exists in the decay $\eta\to\ppp$. This strengthens the
justification of our choice to identify the nonresonant contribution
with the $\ket|1(S)>$ state.
In any case, the arguments given here demonstrate 
that a measurement of the branching fraction
$\BR{D^0\to \pi^0\pi^0\pi^0}$ and, possibly, an analysis of this mode's
Dalitz-plot distribution should shed more light on the role of isospin
symmetry in $D^0$ decays to three-pion final states.

\begin{acknowledgments}
%
This research was supported by INFN, Italy; 
by grant number 2006219 from the United States-Israel Binational
Science Foundation (BSF), Jerusalem, Israel; 
and by the United States National Science Foundation grant number
0457336.
The authors thank Y.~Grossman, J.~Silva, and L.~Wonfenstein for useful
suggestions.

\end{acknowledgments}


\begin{thebibliography}{99}

\bibitem{Aubert:2007ii} 
The \babar\ Collaboration (B.~Aubert {\it et al.}),
  Phys.\ Rev.\ Lett.\  {\bf 99}, 251801 (2007).


\bibitem{ref:zemach}
C.~Zemach, 
  Phys.\ Rev.\ {\bf 133}, B1201 (1964).


\bibitem{ref:pdg06}
Particle Data Group, Y.-M.~Yao {\it et al.},
J.\ Phys.\ G {\bf 33}, 1 (2006).


\bibitem{ref:thesis} K.~Mishra, Ph.D. Thesis, University of Cincinnati, 
SLAC-Report-893, 72-77 (2008).


\bibitem{Rubin:2005py}
  The CLEO Collaboration (P.~Rubin {\it et al.}),
  Phys.\ Rev.\ Lett.\  {\bf 96}, 081802 (2006).


\bibitem{ref:dppp}
The \babar\ Collaboration (B.~Aubert  {\it et al.}), 
  Phys.\ Rev.\ D {\bf 74}, 091102 (2006).



\end{thebibliography}
\end{document}